\documentclass{ws-mpla}
\usepackage{psfig}
\begin{document}

\markboth{Saneesh Sebastian and V C Kuriakose}
{Spectroscopy and Thermodynamics of MSW Black Hole }

\catchline{}{}{}{}{}

\title{Spectroscopy and Thermodynamics of MSW Black Hole}

\author{\footnotesize  Saneesh Sebastian\footnote{
Typeset names in 8 pt Times Roman, uppercase. Use the footnote to 
indicate the present or permanent address of the author.}}

\address{Department of Physics, Cochin University of Science and
Technology, Kochi 682022, India\\
saneeshphys@cusat.ac.in}

\author{V C Kuriakose}

\address{Department of Physics, Cochin University of Science and
Technology, Kochi 682022, India\\
vck@cusat.ac.in
}

\maketitle

\pub{Received (Day Month Year)}{Revised (Day Month Year)}

\begin{abstract}
We study the thermodynamics and spectroscopy of a  2+1 dimensional black 
hole proposed by Mandal et. al\cite{msw}. We put the background space time in Kruskal like co-ordinate and find 
period with respect to Euclidean time. Different 
thermodynamic quantities like entropy, specific heat, temperature etc are obtained. The adiabatic invariant for 
the black hole is found out and quantized using Bohr-Sommerfeld quantization rule. The study shows that the
area spectrum of MSW black hole is equally spaced and the value of spacing is found to be $\hbar$ 

\end{abstract}

\ccode{PACS Nos.: 04.70.Dy, 04.70.-s }
\section{Introduction}
	The black hole solution is one of the basic solutions of Einstein's General Theory of Relativity.
	The black hole solutions basically contain 
a singularity surrounded by an event horizon. Penrose proved mathematically the existence
of singularity in the solutions of Einstein's General Theory of Relativity\cite{rp}. The horizon of black 
hole is mostly like a one sided membrane. It allows the passage of particles and radiation only in one direction.
    The black hole horizon has a thermodynamic behavior which was first showed by Bekenstein\cite{bek2}. He showed that the area 
of the event horizon is a direct measure of black hole entropy. Later it was found that the four laws of black hole dynamics have a 
striking similarity with the four laws of thermodynamics.  This similarity led Hawking to propose that black hole could emit 
particles\cite{hw2}. Gravitational lensing is another important area of black hole physics\cite{ksv1,ksv2,ksv3} and a number of studies exist
in literature on graviational lensing of black holes and naked singularities\cite{ksv4,ksv5,cmc}.

      Bekenstein proposed that  the black hole area is quantized and each quantum has a value $ 8\pi l_{p}^{2}$ where $l_{p}$ 
      is the Planck length and the spectrum is 
equally spaced. Bekenstein's idea is that there is a relation between the area spectrum and the quasinormal modes of 
Black holes\cite{bek}. Shahar Hod\cite{sh} showed that the area spectrum is related to the imaginary part of quasinormal frequency and obtained 
a value $4l_{p}^{2}ln3$. Later Maggiore showed that one should take both real and imaginary parts of the quasinormal modes 
when finding out the area quantum\cite{mg}. Maggiore obtained the same value for minimum area as Bekenstein obtained 

      There are number of works on the thermodynamics and spectroscopy of various types black holes. Periodicity and 
area spectrum of Schwarzschild and Kerr black hole are obtained by Zeng et. al\cite{ze}. Entropy and area spectrum of charged black hole 
using Quasinormal modes are studied by Wei et. al\cite{sw}. Kunstatter\cite{ku} studied the area spectrum of higher diamensional black
hole. Area spectrum of BTZ black hole is found using periodicity method by Larranaga\cite{al} 
	  
	    $2+1$ dimensional black holes acts as a toy model in Einstein gravity. Since the $2+1$ dimensional black hole are in 
reduced dimensions, the Einstein equation is exactly solvable. Both $2+1$ black hole and $3+1$ black hole share a number of properties and hence
it is important to study simple $2+1$ black holes in Einstein gravity. Lower dimensional charged dilaton black hole has
a relation to the lower dimensional string theory so it acts as a tool to study the lower dimensional string theory. The quasinormal modes 
of a charged dilaton black hole is studied by Fernando\cite{sf} and the area spectrum and thermodynamics are also studied\cite{rv}.
Mandal et.al\cite{msw} have obtained a one parameter family of black hole solution to classical dilaton system in two dimensions. This black hole 
is called MSW black hole in this work. Following Zeng\cite{ze} et. al we consider an out going wave which makes periodic motion outside the horizon.
The gravitational system is periodic with respect to the Euclidean time with a period given by the inverse of Hawking temperature. The frequency of 
this out going wave is given by this temperature. Thus the adiabatic invariant term can be found out using surface gravity term. Using the method 
of Kunstatter\cite{ku} we have found the entropy and area spectrum of the MSW black hole.

     In this paper we study the spectroscopy and thermodynamics of MSW black hole. The paper is organized as follows. In section. 2 we discuss the thermodynamic aspects of 2+1 MSW black hole. In sction. 3
we discuss the area spectrum of the black hole. Finally, conclusion is given in section. 4
\section{Thermodynamics of MSW black hole }
The metric for MSW black hole is given by ($c=G=1$ system of units)
\begin{equation}
ds^{2}=-U(r)dt^{2}+\frac{dr^{2}}{U(r)}+\gamma^{2}r d\phi^{2},
\end{equation}
where U(r) is given by\cite{cm}
\begin{equation}
U(r)=8\Lambda\beta r-2M\sqrt{r}.
\end{equation}
This metric has a zero at (ie the horizon) 
\begin{equation}
r_{+}=\frac{M^{2}}{16\Lambda^{2}\beta^{2}},
\end{equation}
 where $\Lambda$ is the cosmological constant, $\beta$ is a constant factor, $M$ is the mass of the black hole. 
 The black hole mass as a function of $r_{+}$ is 
\begin{equation}
 M=4\Lambda\beta \sqrt{r_{+}}.
\end{equation}
\begin{figure}[h]
\centering
\includegraphics{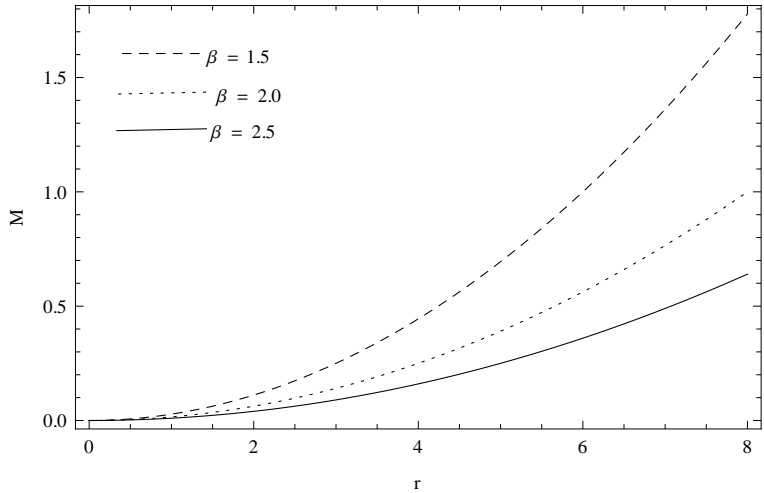}
\caption{Variation of mass with horizon radius for different values of $\beta$}
\end{figure}
The variation of $M$ versus $r_{+}$ is shown in Fig.(1) for the different values of $\beta$ and it has a parabolic nature.  
The entropy $S$ of this $2+1$ dimensional black hole can be obtained as 
\begin{equation}
 S=\frac{A}{4}=\frac{\pi r_{+}}{2}.
\end{equation}
 Here A is the area, since it is a $2+1$ black hole area implies the circumference of the horizon.
 Since $r_{+}$ can be expressed in terms of $M$, we can also express the 
entropy in terms of black hole mass, as  
\begin{equation}
 S=\frac{\pi M^{2}}{32\Lambda^{2}\beta^{2}}.
\end{equation}
\begin{figure}[h]
 \centering
\includegraphics{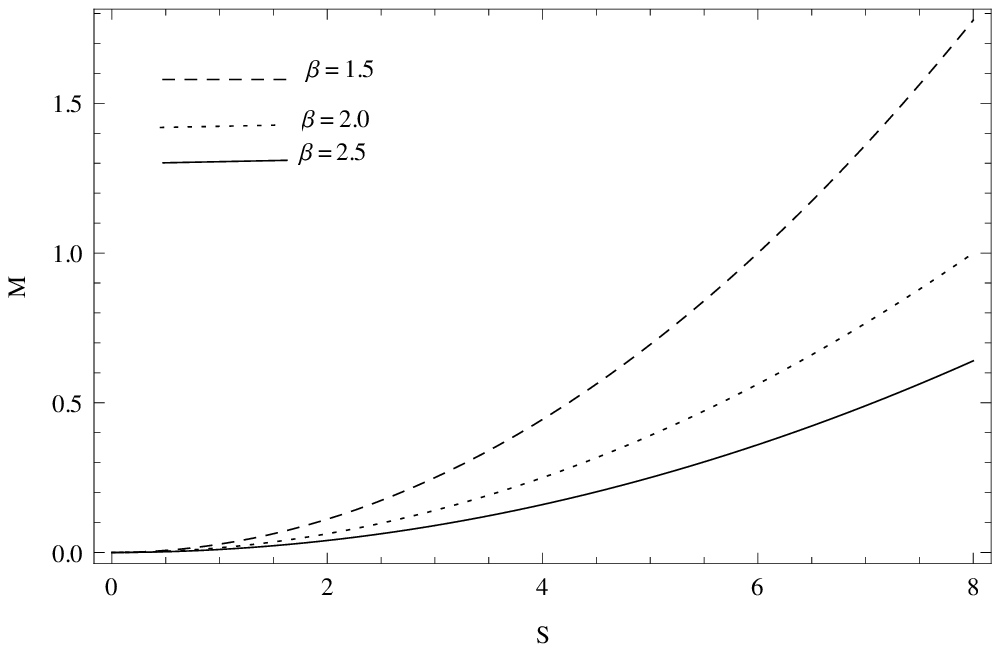}
\caption{variation of entropy with mass for diffrent values of $\beta$}
\end{figure}
Fig.(2) is a plot between entropy and mass and it shows a parabolic nature. 
The black hole temperature can now be obtained since entropy-mass relation is 
known. The general expression for temperature T is 
\begin{equation}
 T=(\frac{\partial M}{\partial S}).
\end{equation}
The heat capacity is given by the general expression 
\begin{equation}
 C=T(\frac{\partial S}{\partial T}).
\end{equation}
Eq.(5) can be inverted such that we get $r_{+}$ in terms of $S$ as
\begin{equation}
 r_{+}=\frac{2S}{\pi}.
\end{equation}
From Eq.(7) we can find out the temperature T as 
\begin{equation}
 T=\frac{4\Lambda\beta}{\pi \sqrt{r_{+}}}.
\end{equation}
\begin{figure}[h]
 \centering
\includegraphics{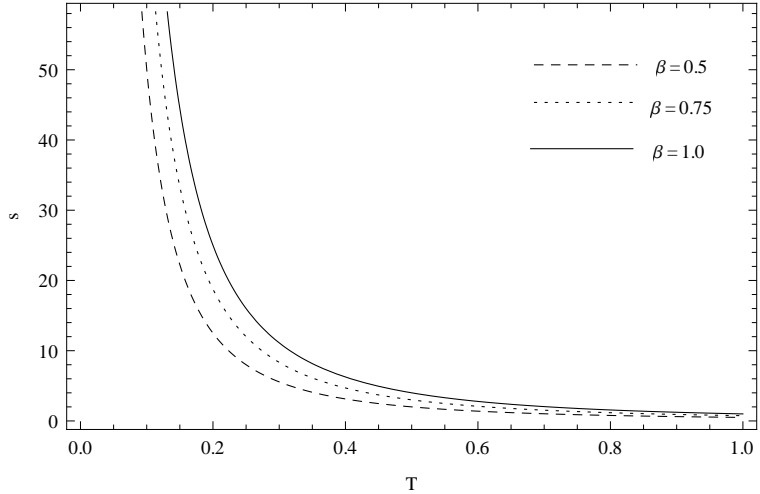}
\caption{variation of entropy with  temperature for different values of$\beta$}
\end{figure}
Fig.(3) shows variation of entropy with temperature. Entropy increases 
rapidly as temperature decreases.
The specific heat of black hole can be obtained from Eq.(8). 
The specific heat is obtained as
\begin{equation}
 C=-\pi r_{+}=-2S.
\end{equation}
The specific heat is negative as usual and is twice the entropy of the black hole. The negative heat capacity indicates that 
the black hole is thermodynamically unstable. For a large number of black holes heat capacity is negative showing that black holes are  
thermodynamically unstesble.
\section{Spectroscopy of MSW black holes}
In this section we study the spectra of MSW black hole using adiabatic invariant method\cite{qy,qq}. We find the area spectrum by evaluating 
the adiabatic invariant integral which varies very slowly compared to the external perturbations. 
The adiabatic invariant is quantized using Bohr-Sommerfeld quantization rule. The first law of black hole thermodynamics says that 
\begin{equation}
 dM=\frac{T_{H}dA}{4}.
\end{equation}
We utilize the period of the motion of out going wave, which is shown to be related to the vibrational spectrum of the perturbed black hole, to quantize 
the area of the MSW black hole. It is well known that the gravity system is periodic with respect to the Euclidean time in Kruskal coordinate.
Particles in this periodic gravitating system also own a period which is inverse of Hawking temperature. To find out area spectrum by periodicity method 
we use the Klein-Gordon equation given by

\begin{equation}
 g^{\mu\nu}\partial_{\mu}\partial_{\nu}\Phi-\frac{M^{2}}{\hbar^{2}}\Phi=0.
\end{equation}
We define an action $A$ such that $\Phi$ is related to the action as
\begin{equation}
 \Phi=exp[\frac{i}{\hbar}A(t,r)].
\end{equation}
Adopting the wave equation ansatz for the scalar field we get the solution of wave equation. We can also obtain the solution from the 
Hamilton-Jacobi equation, 

\begin{equation}
 g^{\mu\nu}\partial_{\mu}A\partial_{\nu}A+M^{2}A=0.
\end{equation}
The action $A(t,r)$ can be decomposed into the functions of $r$ and $t$ and be written as\cite{ze,dy}
\begin{equation}
 A(t,r)=-Et+W(r),
\end{equation}
 where $E$ is the energy of the emitted particle measured by an observer at infinity. $W(r)$ in the action can be written in near horizon as\cite{dy}  
\begin{equation}
 W(r)=\frac{i\pi E}{f'(r_{+})},
\end{equation}
where we consider the out going wave near the horizon. In this case, it is obivious that the wave function $\Phi$ outside the horizon can be expressed 
in the form 
\begin{equation}
 \Phi=exp[-\frac{iEt}{\hbar}]\Psi(r_{+})
\end{equation}
and $\Psi(r_{+})$ is given by
\begin{equation}
 \Psi(r_{+})=exp[\frac{\pi E}{hf'(r_{+})}],
\end{equation}
From Eq.(18) we can find that $\Phi$ is periodic with a period 
\begin{equation}
 \tau=\frac{2\pi\hbar}{E}.
\end{equation}
The gravitating system is periodic in Kruskal coordinate in Euclidean time and the moving particle also gets a periodic motion 
in this periodic gravity. The period is related inversely to the Hawking temperature as
\begin{equation}
 \tau=\frac{2\pi}{\kappa_{r}}=\frac{\hbar}{T_{H}},
\end{equation}
and hence, the Hawking temperature is given by
\begin{equation}
 T_{H}=\frac{\hbar\kappa_{r}}{2\pi}. 
\end{equation}
 In the  near horizon 
\begin{equation}
 U(r)=-M+4\Lambda\beta\sqrt{r_{+}}.
\end{equation}
As $U(r)\rightarrow 0$, the change in the mass $dM$, becomes
\begin{equation}
 dM=2\Lambda\beta \frac{dr_{+}}{\sqrt{r_{+}}}.
\end{equation}
The adiabatic invariant\cite{qq} is defined as 
\begin{equation}
 I=\int{\frac{2\pi dM}{\kappa_{r_{+}}}},
\end{equation}
where $ \kappa_{r_{+}}$ is the surface gravity and is given by 
\begin{equation}
 \kappa_{r_{+}}=\frac{dU(r)}{dr}|_{r=r_{+}}.
\end{equation}
From (23)
\begin{equation}
 \frac{dU(r)}{dr}|_{r=r_{+}}=\frac{4\Lambda\beta}{2\sqrt{r_{+}}},
\end{equation}
and using Eq.(26), we can have 
\begin{equation}
 \kappa_{r_{+}}=\frac{2\Lambda\beta}{\sqrt{r_{+}}}
\end{equation}
Substituting Eqs.(28) and (24) in Eq.(25) and applying Bohr-Sommerfeld quantization rule, we get
\begin{equation}
 2\pi r_{+}=A_{n}=n\hbar.
\end{equation}
and $ \Delta A$ is given by 
\begin{equation}
 \Delta A=A_{n+1}-A_{n}=\hbar
\end{equation}
Thus we see that the circumference spectrum of MSW black hole is discrete and the spacing is equidistant. For this system the circumference spectrum 
is indepedent of black hole parameters. Since the entropy of this system is related to circumference spectrum, it is also quantized.
\section{Conclusion}
The thermodynamics of 2+1 dimensional MSW black holes are studied and area spectrum is obtained using adiabatic invariant method.
We have obtained the heat capacity, temperature and mass of MSW black hole. (The different quantities are plotted and studied their behavior)
The mass of the black hole is plotted for various values of horizon radius with different parameter $\beta$ and the graph shows that the horizon 
radius increases quadratically with mass, the same behavior is shown for mass versus entropy. The variation of temperature with entropy is 
also plotted and the graph shows that the entropy increases very rapidly with decreasing temperature. The heat capacity is found to be negative 
as is the case for most black holes. This black hole does not show any phase transition. The period of Euclidean time is found out 
and adiabatic invariant is calculated. Using Bohr-Sommerfeld quantization rule we have obtained the area spectrum, here the circumference spectrum
and is found to be quantized and equally spaced. 
\section*{Acknowledgement}
SS wishes to thank CSIR, New Delhi for financial support under CSIR-SRF scheme. VCK wishes to acknowledge Associateship of IUCAA, Pune, India

\end{document}